\documentclass[10pt,a4paper,twocolumn]{article}
\usepackage[utf8]{inputenc}
\usepackage{amsmath}
\usepackage{amsfonts}
\usepackage{amssymb}
\usepackage{graphicx}
\usepackage{url}
\usepackage[margin=2.5cm]{geometry}

\def \calE {\mathcal{E}}
\def \calF {\mathcal{F}}
\def \calG {\mathcal{G}}
\def \calN {\mathcal{N}}
\def \calP {\mathcal{P}}
\def \calR {\mathcal{R}}

\title{Evolution of Communities with Focus on Stability}
\author{Carlos Sarraute  \\
Grandata Labs \\
Buenos Aires, Argentina \\
\url{charles@grandata.com}
\and Gervasio Calderon \\
Grandata Labs \\
Buenos Aires, Argentina \\
\url{gerva@grandata.com}
}
\date{}

\begin{document}
\maketitle
\thispagestyle{empty}
\pagestyle{empty}

\section{Introduction}

The detection of communities is an important tool used
to analyze the social graph of mobile phone users.
Within each community, customers are susceptible of attracting new ones,
retaining old ones and/or accepting new products or services through 
the leverage of mutual influences \cite{wu2009group}.
The communities of users are smaller units, easier to grasp,
and allow for example the computation of
role analysis -- based on the centrality of an actor within his community.

The problem of finding communities in static graphs has been widely studied
(see \cite{fortunato2010community} for a survey).
However, from the point of view of a telecom analyst,
to be really useful, the detected communities must evolve as the 
social graph of communications changes over time  --
for example, in order to perform marketing actions on communities and
track the results of those actions over time.
Additionally the behaviors of communities of users over time
can be used to predict future activity that interests the
telecom operators, such as subscriber churn or handset adoption \cite{greene2010tracking}.
Similary group evolution can provide insights for designing strategies, 
such as the early warning of group churn.

Stability is a crucial issue: the analysis performed on a given community will
be lost, if the analyst cannot keep track of this community in the following time steps.
This is the particular use case that we tackle in this paper: 
tracking the evolution of communities in dynamic scenarios with focus on stability.

We propose two modifications to a widely used static community detection algorithm.
We then describe experiments to study the stability and quality of the resulting partitions
on real-world social networks, represented by monthly call graphs for millions
of subscribers.

\section{Data Sources} \label{sec:data_sources}

Our raw data source is anonymized traffic information from a mobile operator.
The analyzed information ranges from January 2012 to January 2013,
and contains for each communication the origin, target, 
date and time of the call or sms, and duration in the case of calls.

For each month $T$, we construct a social graph $\calG_{T} = < \calN_{T}, \calE_{T} > $.
This graph is based on the aggregation of the traffic of several months,
more concretely $\calG_{T}$ depends on the traffic of three months: $T$, $T-1$ and $T-2$.
The raw aggregation of the calls and messages gives a first graph
with around 92~M (million) nodes and 565~M edges (on a typical month).
The voice communications contribute 413~M edges and the messages contribute 296~M edges
to this graph.

We then perform a symmetrization of the graph, keeping only the edges $(A,B)$
whenever there are communications from $A$ to $B$ and from $B$ to $A$.
This new graph has around 56~M nodes and 133~M (undirected) edges,
and represents stronger social interactions between nodes.
Additionally we filter nodes with high degree (i.e. degree greater than 200)
since we are interested in the communications between people (and not call centers or
platform numbers).

\section{Dynamic Louvain Method}

Our first experiment to detect evolving communities was to run the 
original Louvain algorithm \cite{blondel2008fast} on the 
graphs at time $T$ and $T+1$, and compare the two partitions,
method that resulted very unstable.
Our second experiment was to run the Louvain algorithm modified
by Aynaud and Guillaume \cite{aynaud2010static} 
to obtain a more stable evolution.
As we show in Section~\ref{sec:results} the results were still 
unsatisfying in terms of stability.

In our use case (e.g. telecom analysts performing actions on the communities), 
the stability of the partition is our main concern.
With this goal in mind, we propose two modifications to the Louvain method,
that give the partition at the previous time step a sort of ``momentum",
and make it more suitable to track communities in dynamic graphs.

Before describing them, we introduce some notations.
As stated in the previous section, we consider snapshots of the social graph
constructed at discrete time steps (in our case every month).
Let $\calG_{T} = < \calN_{T}, \calE_{T} >$ be a graph that has already
been analyzed and partitioned in communities. Let $\Gamma = <C_1, \ldots, C_R>$
be such partition in $R$ communities.
Given a new graph $\calG_{T+1} = < \calN_{T+1}, \calE_{T+1} >$
our objective is to find a partition of $\calG_{T+1}$ which
is stable respect to $\Gamma$.

The first idea is to have a set of \emph{fixed nodes} $\calF$.
Let $\calR = \calN_{T} \cap \calN_{T+1}$  be the set of nodes that remain from 
time $T$ to $T+1$.
The set $\calF$ is a subset of $\calR$, whose  
nodes are assigned to the community they had at time $T$.
In other words, noting $\gamma$ the function that assigns a community to each node, 
we require: 
$ \gamma_{T+1}(x) = \gamma_{T}(x) \; \forall x \in \calF $.

We experimented with different distributions of the fixed nodes,
ranging from no fixed nodes ($\calF = \emptyset$) to all the remaining nodes 
($\calF = \calR$). 
For the experimental results, we used a parameter $p$
that represents the probability that a node belongs to $\calF$
(i.e. $| \calF | = p \cdot | \calR | $).

The second idea is to add a probability $q$ of ``preferential attachment"
to pre-existing communities.
With probability $q$, the new nodes will prefer to attach to a community existing at time $T$
instead of attaching to a community formed at time $T+1$.
We give the details below.

The Louvain Method \cite{blondel2008fast} is a hierarchical greedy algorithm, composed of two phases.
During phase 1, nodes are considered one by one, and each one is placed
in the neighboring community (including its own community)
that maximizes the modularity gain.
This phase is repeated until no node is moved (that is when the decomposition reaches
a local maximum).
Then phase 2 consists in building the graph between the communities obtained during phase 1.
Then the algorithm starts phase 1 again with the new graph, in the next hierarchical
level of execution, and continues until the modularity does not improve anymore.

We construct a set $\calP \subseteq \calN_{T+1}$
such that $|\calP| = q \cdot | \calN_{T+1} |$.
For every node $x$, we consider its neighbors that belong to a community existing 
at time $T$, that is the set 
$ A(x) = \{ z \in \calN_{T+1} \, | \, (x,z) \in \calE_{T+1} \wedge \gamma_{T+1}(z) \in \Gamma_T  \} $.
During phase 1 of the first iteration of the algorithm (i.e. during the first
hierarchical level of execution), the inner loop is modified.
For all node $x \in \calN_{T+1}$, if $x \in \calP$ and $A(x) \neq \emptyset$ then place $x$ in the community
of $A(x)$ which maximizes the modularity gain
(whereas if $A(x) = \emptyset$ proceed as usual).

\section{Experimental Results} \label{sec:results}

\begin{figure}[t]
\vspace{-0.4cm}
\centerline{\includegraphics[width=0.50 \textwidth]{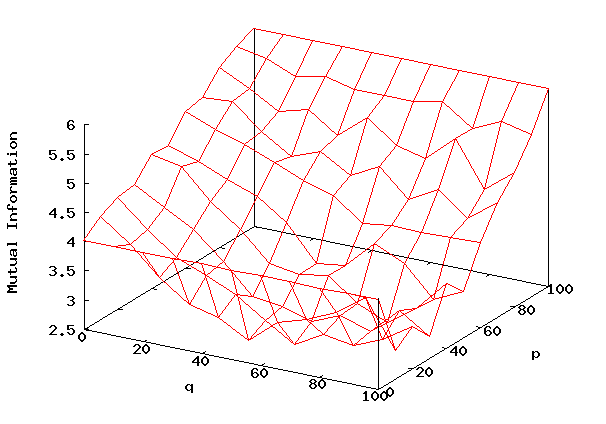}}
\vspace{-0.2cm}
  \caption{Mutual Information  as a function of $p$ and $q$ (expressed as percentages).}
  \label{fig:mutual_information}
\end{figure}

\begin{figure}[t]
\vspace{-0.4cm}
\centerline{\includegraphics[width=0.50 \textwidth]{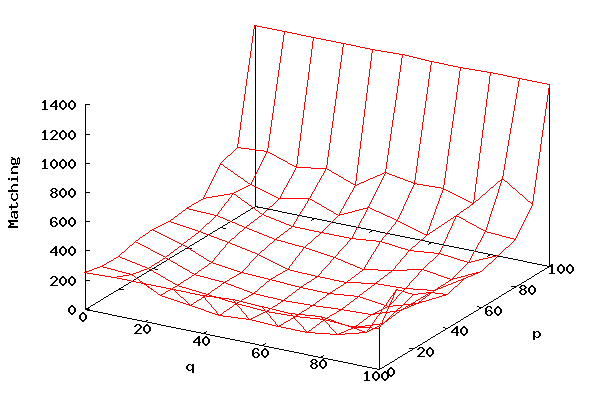}}
\vspace{-0.2cm}
  \caption{Matching communities.}
  \label{fig:matching}
\end{figure}

\begin{figure}[t]
\vspace{-0.5cm}
\centerline{\includegraphics[width=0.50 \textwidth]{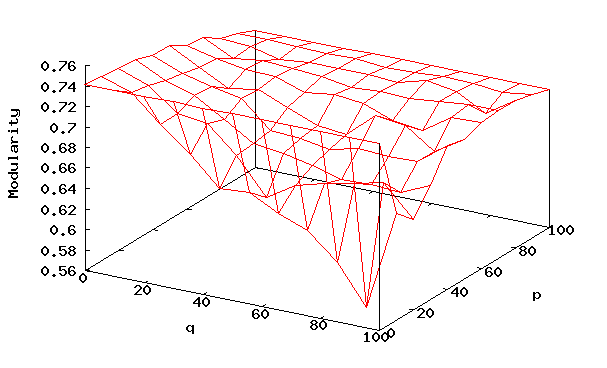}}
\vspace{-0.2cm}
  \caption{Modularity.}
  \label{fig:modularity}
\end{figure}

In our experiments, we computed the social graph 
(constructed as described in Section \ref{sec:data_sources}).
Since we are interested in the real-world application of our method,
we preferred to evaluate it on real data.

Given two months $T$ and $T+1$, we calculated a partition in communities
of $\calG_{T}$ using the Louvain Method (with the modification of \cite{aynaud2010static})
that we note $\Gamma = < C_1, \ldots, C_R > $;
and a partition of $\calG_{T+1}$ using our dynamic version of the Louvain Method,
with different values of the parameters $p$ and $q$. 
Let $\Gamma' = < C'_1, \ldots, C'_S > $ be the partition of $\calG_{T+1}$.
We are interested in comparing $\Gamma$ and $\Gamma'$ in terms of stability
and quality of the partition. To this end, we measure:
 (i) the mutual information between the two partitions;
 (ii) the number of matching communities (i.e. such that the proportion 
 of nodes in common is greater than a parameter $r$);
 (iii) the final modularity of $\Gamma'$ (as defined in \cite{newman2004finding}).

The number of matching communities is computed as follows: for each
community $C'_j \in \Gamma'$, we evaluate whether there is a community
$C_i \in \Gamma$ such that $ | C_i \cap C'_j | > r \cdot | C_i | $
and $ | C_i \cap C'_j | > r \cdot | C'_j | $, where $r$ is a fixed
parameter verifying $r > 0.50$ (for instance we used $r = 0.51$).
In that case, we say that $C'_j$ matches $C_i$. 
The matching communities are of particular interest,
because $C'_j$ can be considered as the evolution of $C_i$ (although the community
may have grown or shrank) and can be individually followed by a human analyst.

The mutual information for two partitions of communities
(see \cite{fenn2012dynamical,greene2010tracking} for definitions\footnote{Since
nodes can change between time $T$ and $T+1$, we only consider the intersection
$\calN_{T} \cap \calN_{T+1}$ for the mutual information computation.}) 
is computed as: 
$$
MI(\Gamma,\Gamma') = \sum_{i=1}^{R} \sum_{j=1}^{S} P(C_i,C'_j) 
\log \frac{ P(C_i,C'_j) }{ P(C_i) \cdot P(C'_j) } .
$$

To analyze the effect of $p$ and $q$, we made those parameters vary from $0$ to $1$.
The baseline, for $p=0$ and $q=0$, corresponds to the Louvain Method
with the modifications of \cite{aynaud2010static}.

Fig.~\ref{fig:mutual_information} shows the effect on the mutual information
between the two partitions. We can clearly observe that the mutual information
increases as $p$ increases, and reaches its maximal values at $p = 100\%$.
The effect of varying $q$ is not so clear, since it produces fluctuations
of the mutual information without a marked tendency.

Fig.~\ref{fig:matching} shows the number of matching communities (according to our
criterion). In this graph we see that the number of matching communities 
increases dramatically when $p$ approaches $100\%$.
The effect of varying $q$ is again not clearly marked, although the increase
of $q$ produces higher matching communities for smaller values of $p$.

Fig.~\ref{fig:modularity} shows the effect on the modularity of the new partition.
We can observe that the modularity decreases slightly as $p$ increases
for small values of $q$. For greater values of $q$ (closer to $100\%$),
varying $p$ produces fluctuations with a decreasing tendency.

As a conclusion, we can see that increasing the probability $p$ of fixed nodes
has a clear effect on increasing the mutual information between the two partitions,
and the number of matching communities. The trade-off with quality is good,
since the decrease in modularity is relatively low.

On the other side, increasing the probability $q$ of preferential attachment
to pre-existing communities has not a clear effect on mutual information or matching communities.
It does not seem advisable to use this second modification for generating evolving communities.

\section{Conclusion and Future Steps}

The  detection of evolving communities is a subject that still requires further
study from the scientific community.
We propose here a practical approach for a particular version of this problem
where the focus is on stability.
The introduction of fixed nodes (with probability $p$)
increases significantly the stability of successive partitions,
at the cost of a slight decrease in the final modularity of each partition.

As future steps of this research, we plan to:
(i) study the evolution of communities 
with finer grain, using smaller time steps;
(ii) evaluate the proposed method on publicly available datasets,
to facilitate the comparison of our results;
(iii) refine the matching criteria, and consider additional events
in the evolution of dynamic communities (such as 
birth, death, merging, splitting, expansion and contraction \cite{greene2010tracking}).


\bibliographystyle{unsrt}

\bibliography{sna}

\end{document}